# Graphene Induced Surface Reconstruction of Cu

*Jifa Tian[1,2], Helin Cao[1,2], Wei Wu[3], Qingkai Yu[4], Nathan P. Guisinger[5,\*], Yong P. Chen[1,2,6,\*]*

1. Department of Physics, Purdue University, West Lafayette, Indiana 47907

2. Birck Nanotechnology Center, Purdue University, West Lafayette, Indiana 47907

3. Center for Advanced Materials and Department of Electrical and Computer Engineering, University of Houston, Houston, Texas 77204

4. Ingram School of Engineering, and Materials Science, Engineering and Commercialization Program, Texas State University, San Marcos, Texas 78666

5. Center for Nanoscale Materials, Argonne National Laboratory, Argonne, Illinois 60439

6. School of Electrical and Computer Engineering, Purdue University, West Lafayette, Indiana 47907

*To whom correspondence should be addressed: E-mail: nguisinger@anl.gov and yongchen@purdue.edu

**ABSTRACT** An atomic-scale study utilizing scanning tunneling microscopy (STM) in ultrahigh vacuum (UHV) is performed on large single crystalline graphene grains synthesized on Cu foil by a chemical vapor deposition (CVD) method. After thermal annealing, we observe the presence of periodic surface depressions (stripe patterns) that exhibit long-range order formed in the area of Cu covered by graphene. We suggest that the observed stripe pattern is a Cu surface reconstruction formed by partial dislocations (which appeared to be stair-rod-like) resulting from the strain induced by the graphene overlayer. In addition, these graphene grains are shown to be more decoupled from the Cu substrate compared to previously studied grains that exhibited Moiré patterns.

KEYWORDS  Graphene, Cu, STM, Strain, Surface Reconstruction, Dislocation.

MANUSCRIPT TEXT Graphene, one atomic layer of carbon organized in a honeycomb lattice, has exhibited spectacular electronic properties and stimulated intense research interests for applications in nanoelectronics.[1,2] Since the pioneering work on exfoliated single layer graphene, numerous efforts have been made to advance both the techniques of growing graphene and the investigations of graphene properties.[2-10] The successful synthesis of large-area graphene films[5-10] has accelerated the potential applications of graphene in nanoelectronics and other technological areas. In the past few years, the CVD growth technique has been applied to synthesize large-scale graphene on various metal surfaces. By this technique, a hydrocarbon gas (such as $CH_4$) is passed at high temperature over metal substrates, such as Cu[7,8,9], Ni[6,10,11], Ru[12-14] and Ir[15-17], resulting in the formation of graphene layers on the surface. Among them, Cu foil has become the most popular substrate due to its low cost, and the fact that graphene films grown on it are predominantly monolayer and can be easily transferred to other substrates[7-9]. On the other hand, grain boundaries that form in the continuous films limit their carrier mobility and induce carrier scattering that could degrade the quality of graphene devices.[18-20] Recently, we synthesized and characterized hexagonally-shaped single crystalline graphene grains on Cu foils grown by CVD techniques [18,19]. Here, we further study the interaction between graphene grains and underlying Cu foil substrate at the atomic-scale using STM. Such a study may also help us understand

the graphene growth mechanism and improve the quality of the CVD graphene films.

STM has been widely used to probe the atomic-scale structure and electronic properties (such as local density of states, LDOS) of graphene films grown on silicon carbide[5,21], metal substrates[12,14,22,23], and exfoliated graphene on $SiO_2$/Si[24-26]. In our work, the STM has been utilized to study graphene grains grown on polycrystalline Cu foil. Particular efforts are dedicated to study the interaction between graphene and underlying Cu substrate. The STM experiments are performed on the as-grown grains transferred into UHV followed by annealing at 400 °C. Our investigation reveals the presence of well-ordered stripe patterns with ~5nm periodicity confined to the regions of large-area graphene grains. These stripes are interpreted as a reconstruction of Cu surface due to partial dislocations formed beneath the graphene in the surface of the Cu to relax the tensile strain induced by the graphene overlayer. Furthermore, we find that these graphene grains are more decoupled from the substrate compared to those studied in several previous reports[22,27] (where the graphene domains are much smaller than ours, did not induce dislocations in the underlying Cu, and exhibited Moiré patterns).

The graphene samples used in this work are synthesized on polycrystalline Cu foils (25-μm-thick, 99.8%, Alfa Aesar) by ambient CVD ($CH_4$ as carbon stock), using similar procedures as in our previous publications[18,28]. The growth is carried out at 1050 ºC and stopped before the graphene grains would have merged with each other to form a globally continuous[9,29] (but polycrystalline) graphene film. Then, the sample is rapidly cooled to room temperature under the protection of Ar and $H_2$. Finally, the samples are removed from the CVD tube furnace (thus exposed to air) before being transferred into the UHV STM chamber.

The STM measurements are carried out at room temperature in an Omicron STM with a base pressure < $10^{-11}$ mbar, with electrochemically etched STM tips made of tungsten or platinum/iridium alloy. After the sample is moved into the UHV chamber, it is annealed at 400 °C for 48 h. We perform both STM topography as well as scanning tunneling spectroscopy (STS) measurements. The STS and differential conductance (*dI/dV*) are measured with lock-in detection by applying a small modulation (30 mV rms in amplitude) to the tunneling voltage at ~10 KHz.

The typical shape of a graphene grain is hexagonal[18] and in some instances we observe merged multiple hexagonal grains (Fig. S1a, showing an SEM image after taking out of CVD furnace, but before UHV thermal annealing). A morphological difference between bare Cu and graphene grains can be clearly observed (see also Fig. S1b, showing the STM image of a typical unannealed sample). A significant source of the observed roughness on graphene is attributed to the steps on the underlying Cu substrate[18], giving rise to the line texture seen in Figs. S1a,b. These line textures are nonuniform with spacing over hundreds of nanometers and height differences ~10 nm[18]. Such steps are nearly unobservable on bare Cu surface not covered (and protected) by graphene, due to the formation of a thin layer of native copper oxide.[18,27,30] These results reveal that graphene appears to be an excellent impermeable barrier against oxidation or other forms of surface chemistry on Cu. Our previous STM study demonstrated that high temperature annealing can help remove such Cu oxide.[27]

Once introduced into the UHV STM system, the sample is annealed at 400 °C for 48 h, following which large single crystalline graphene grains are located, as illustrated in the STM topographic image of Fig. 1a. On the Cu surface outside graphene, the oxide can be partially removed by this thermal annealing process. From the STM topographic image (Figs. 1a, compared to the unannealed samples, Fig. S1a,b), we can see that the post-annealing Cu surface outside graphene becomes much smoother and shows up the Cu atomic steps. Most strikingly, we find that the surface morphology *within* the graphene grain has also changed dramatically after the annealing, where a long range periodic stripe pattern (appearing as many periodic dark lines largely parallel to one of the edges of the graphene grain in the topographic image Fig. 1a) is observed. Fig. 1b further shows a few such dark lines in a zoomed-in atomically resolved topographic image taken from the area indicated by the blue dashed box in Fig. 1a. There are no apparent distortions, defects or electronic scattering within the graphene lattice at the dark lines (see also Fig. S2). This stripe pattern, which will be interpreted as a graphene-induced Cu surface reconstruction, is the main finding of this paper.

Figure 1c shows another example of such a stripe pattern observed in the graphene region. The stripes again appear in the topographic image as uniformly spaced dark lines, corresponding to depressions in the surface. Two additional interesting features are observed in this case. One is that within this graphene region the orientation of the stripes is found to change by ~18°, as seen in Fig. 1c from the upper-right to the lower-left of the image. Taking a Fourier transform (FT) of this image, we can see that the stripe pattern gives rise to two sets of spots (indicated by green and red circles in Fig. S3, respectively). The angle between the two sets of spots corresponds to the orientation change (~18°) of the stripe pattern. Another interesting observation is the two new dark lines initiating from the locations highlighted by the dashed black and red circles in Fig. 1c near the transition region between the two orientations. Fig. 1d shows a high magnification STM image for the stripe pattern taken from the region highlighted by the dashed black circle in Fig. 1c. Importantly, it can be seen that the graphene lattice remain coherent (single crystalline) within the entire region, even around where the orientation of the stripes changes or the new dark line initiates.

An atomically resolved STM image and the corresponding height profile of a zoomed-in region containing several such well-ordered stripes (dark lines, indicating surface depressions) are shown in Figs. 2a,b, respectively. It can be seen that the spacings of these dark lines are uniform, with a periodicity ~ 5nm. The depth of the depression measured on graphene surface is ~ 0.6 Å (Fig.2b). Figure 2c gives the three dimensional (3D) rendition of a high resolution STM topographic image acquired around one of such lines. We can again clearly see the graphene lattice highlighted by the blue hexagons remains intact crossing the dark line.

While in principle both wrinkles (ripples) on graphene [31,32] as well as Moiré patterns[27,33] between graphene and Cu could give rise to line patterns, neither of those are consistent with the features of the stripe pattern we have observed. The graphene wrinkles or ripples typically appear as surface protrusions [31,32], not depressions. They are also more irregularly-spaced, with much larger sizes and separations than our dark lines [31,32]. The periodicity of the Moiré pattern is sensitive to its orientation [27,33], whereas our stripe pattern maintains its periodicity even with an orientation change (Fig. 1c). Furthermore, the dark lines in our stripe pattern are not perfectly straight and can also show occasional bending or curvatures (as seen in Figs. 1a, 1c, 2a), unlike the highly straight lines that would be produced by a lattice superstructure from the Moiré pattern[27,33]. Therefore we conclude that the dark lines (which appear to minimally perturb the graphene lattice) of our stripe pattern most likely arise from features (in this case depressions) formed on the Cu surface *underneath* the graphene. Since they only appear under graphene after appropriate thermal annealing, they represent a graphene-induced reconstruction of Cu surface, and we speculate that they result from partial dislocations due to the strain between graphene and Cu, as discussed below.

It is well known that the Cu surface and adsorbate-covered Cu surface can reconstruct.[34-38] Previous studies on Cu covered by adatoms further showed that such surface reconstruction can be quite sensitive to thermal annealing as well as adatom coverage.[34,35] The reconstruction feature sizes and spacings are on the similar orders of magnitudes as those in our observation. In addition, lattice reconstructions due to formation of partial dislocations[39] to release the strain are common in many metal surfaces or heteroepitaxial thin films, as have been reported in numerous systems such as Au(111),[40,41] Ni/Cu(001),[42] Ni/Ir(100),[39] Co/Ir(100),[39] Cu/Ru(0001)[43] etc. In our case, although the CVD growth of graphene on Cu foil may not be epitaxial[18], strain can still be induced after growth at high temperature followed by cooling down to room temperature, during which the graphene lattice can expand (due to its negative coefficient of thermal expansion[6-8,44]) while Cu shrinks, giving rise to a tensile strain on Cu. During the subsequent thermal annealing in UHV, the top few atomic layers of Cu underneath graphene may become unstable and start to reconstruct, forming (periodic) depressions on the Cu surface thus releasing the tensile strain. A possible mechanism for such reconstruction of the top few atomic layers of Cu underneath graphene is through the formation of partial dislocations. The partial dislocations can give rise to the long-range ordered depressions/stripes in Cu shown as the dark lines in the STM topographic images. We note that such dark lines (depressions) are reminiscent of stair-rod dislocations (a common form of surface misfit partial dislocations) previously found in heteroepitaxial thin films[39]. It is also worth noting that in our case the partial dislocations are formed in the surface layers of the bulk substrate (Cu) instead of the overlayer (graphene). Indeed the graphene lattice appears to be intact with no signs of reconstruction in our STM images (Figs. 1b,d, 2a,c, and S2). This is a novel situation and very different from the case observed in the growth of many heteroepitaxial thin films[39, 42,43] where the dislocations are typically formed in the overlayer (heteroepitaxial thin films). We believe this difference reflects the high mechanical strength and stability of graphene lattice[45], making the reconstruction happen in Cu rather than graphene during thermal annealing.

Figure 2d displays a proposed schematic model for an example of the stair-rod-like[39] dislocation in Cu under graphene. The model is presented for a Cu surface orientation of (100). Even though we do not have a direct measurement of the local Cu surface orientation under the stripe pattern, (100) has been found to be the predominant surface orientation (with no other orientations detected) for our Cu foil from our X-ray diffraction measurements,[27] as well as the linear Moiré pattern[27] (see also Fig. S4 and discussions below) observed from smaller and further annealed graphene grains on the same Cu foil. The principal mechanism for the formation of such a stair-rod dislocation[39] can be understood as the interaction of two partial dislocations taking place at the Cu (111) and ($1\bar{1}\bar{1}$) faces (resulting from the refaceting of the Cu (100) surface under tensile strain). The result of the interaction of the two dislocations, which meet at their apex and eventually form a stair-rod-like dislocation, is displayed in Fig. 2d. We emphasize that the proposed mechanism and microstructure for the observed reconstruction as stair-rod-like dislocations are speculative at this point, and need confirmation by further experimental (including additional probes such as LEED (low energy electron diffraction)) as well as theoretical studies. More work is also needed to better understand the physical conditions for the reconstruction to occur, what sets its orientation and periodicity, and whether qualitatively similar surface reconstructions can occur generically on Cu surfaces with different orientations.

The long-range periodic stripe pattern associated with graphene-induced Cu surface reconstruction in this work has not been seen in previous STM studies on unannealed or low temperature (300 $^o$C) annealed graphene grains on Cu foils[18], or on small graphene grains (on single crystalline Cu[22] or Cu foils[27]) after further annealing or thermal cycling at much higher temperatures than ours. It is possible that the Cu surface reconstruction (partial dislocations) we observed could be dependent on various factors such as the thermal annealing or processing condition (as found to be the case previously for Cu surface with adsorbates[34,35]), the size of graphene grains (as smaller grains may be able to release the strain more easily) and other conditions of Cu substrate (eg., impurities) underneath the graphene grain, etc. In addition, the previous work in Ref. 27 (which studied much-smaller-sized, further annealed graphene grains on the same Cu foil as in our current work) observed clear Moiré patterns (formed between graphene and Cu (100) lattices). No Moiré patterns are observed in our (larger size and less annealed) graphene grains. We believe this absence of Moiré patterns reflects a weaker electronic coupling between our graphene grain and Cu substrate than that in Ref. 27. This is also consistent with the much larger apparent graphene-Cu height difference we observe for our graphene grain than for smaller, further-annealed grains similar to Ref. 27 (Fig. S4).

We have further characterized the electronic properties of our graphene grains using STS spectroscopy and mapping[22]. Fig. 3b is a differential conductance *dI/dV* map (at a sample bias of -200 mV) measured concurrently with the topographic image Fig. 3a and shows a higher LDOS over the graphene grain compared to the outside Cu. This electronic contrast allows us to easily identify the two materials when imaging large areas. We have also performed individual *dI/dV* spectral measurements at 100 points along a line (the black dashed line in Fig. 3b with numbers "1", "2" and "3" highlighting a few representative points). The collection of 100 spectra is shown as a 2D color plot in Fig. 3c. It is worth noting that the transition between the electronic properties of the graphene (between "2" and "3") and Cu regions (between "1" and "2") is very abrupt, while the spectra taken within the same region (either graphene or Cu) are similar. Three individual spectra from this measurement are plotted in Fig. 3d, where the red curve is measured over the Cu surface outside the graphene grain and the black and blue curves inside the graphene grain (measured from locations "A" and "B" in Fig. 3b, on and away from a Cu dislocation line, respectively). There is no significant difference between the *dI/dV* spectra measured from "A" and "B", except for a slightly lower value of *dI/dV* in the spectrum "B" on the positive bias side, affirming that the surface depressions (dark lines) on Cu do not appreciably perturb the electronic properties of the graphene overlayer. The "bump" feature in the negative bias side measured in Cu spectrum (observed to be even more prominent on bare Cu foil more extensively annealed with most of the native oxide removed, Fig. S5) is not seen in the graphene spectrum, again consistent with the relatively weak electronic coupling between our graphene gain and Cu (whereas in a previous study[22] of graphene grains strongly coupled to Cu and exhibiting Moiré patterns, the graphene can be nearly "electronically" transparent with similar *dI/dV* spectrum as Cu, as shown in the inset, which is reproduced from Fig. 3c of Ref. 22).

In summary, we have conducted atomically-resolved topographic and spectroscopic investigations of the graphene grains on Cu foil by STM. After annealing at 400 $^{o}$C for 48 h, the Cu surface underneath the graphene grain forms periodic stripe patterns, which we associate with a reconstruction of the Cu surface due to the formation of partial dislocations (such as stair-rod-like dislocations) to release the strain induced by graphene. The graphene lattice remains intact without defects when it crosses the dislocations underneath. No Moiré patterns are observed in graphene and the coupling between our graphene grain and the Cu substrate appears to be relatively weak. Our findings may shed light on the interaction between Cu and graphene as well as the growth mechanism of graphene on Cu foils.

ACKNOWLEDGMENT This work was performed under the auspices of Argonne National Laboratory (ANL) Center for Nanoscale Materials (CNM) User Research Program (Proposal ID 998), and partially supported by NSF, DHS and NRI-MIND center. The user facilities at ANL's CNM are supported by the U.S. Department of Energy, Office of Science, Office of Basic Energy Sciences, under Contract No. DE-AC02-06CH11357." NPG acknowledges DOE "SISGR" Contract No. DE-FG02-09ER16109. The authors also thank Prof. C. K. Shih for valuable discussions.

FIGURE CAPTIONS

**Figure 1.** The STM images of graphene grains on Cu foil: (**a**) An STM topographic image of a graphene grain after annealing at 400 °C for 48 h. The scale bar is 40 nm. Measurement conditions: tunneling current I = 100 pA, sample-tip bias voltage V = -200 mV. (**b**) The zoomed-in atomic-resolved STM topographic image taken from the blue dashed box in (a) (measurement conditions: I = 20 nA, V = -200 mV). The scale bar is 2 nm. (**c**) Another low magnification STM topographic image of the stripe patterns and graphene grain (measurement conditions: I = 1 nA, V = -200 mV). The dotted circles (in black and red) indicate locations where new lines are initiated in the stripe patterns. The scale bar is 50 nm. (**d**) The zoomed-in STM topographic image from the area indicated by a dotted black circle in (c). The scale bar is 5 nm (I = 20 nA, V = -200 mV).

**Figure 2.** Characterization and interpretation of the observed stripe pattern as reconstruction due to partial dislocations on Cu surface underneath graphene: (**a**) An STM topographic image from the area indicated by green dashed box in Fig. 1c, showing surface depressions (interpreted as partial dislocation lines on Cu) under graphene lattice. The scale bar is 6 nm (I = 20 nA, V = -200 mV). (**b**) The height profile along the green dashed line shown in (a). (**c**) Three dimensional (3D) rendition of a high resolution STM topographic image acquired near a typical dislocation line (I = 20 nA, V = -200 mV). (**d**) Schematic model (not to scale) of a stair-rod-like dislocation on Cu (with surface orientation assumed to be (100) in this example) underneath graphene grain.

**Figure 3.** The STM image and scanning tunneling spectroscopy (STS) of a graphene grain on Cu foil. (**a**) STM topographic image showing both graphene and Cu foil. (**b**) Differential conductance *dI/dV* map recorded simultaneously with the topographic image (a). For (a, b), the scale bar is 25 nm (I = 1 nA, V = -200 mV). (**c**) *dI/dV* spectra recorded at 100 points in equal distance along the black line in (b) with numbers "1", "2" and "3" indicating three representative locations labeled in (b). (**d**) Representative *dI/dV* spectra recorded inside (at a two representative locations "A" and "B" in (b)) and outside a graphene grain. Inset (reproduced from Ref. 22) shows *dI/dV* spectra in a previously study[22] of nm-sized graphene islands on Cu (111).

SYNOPSIS TOC

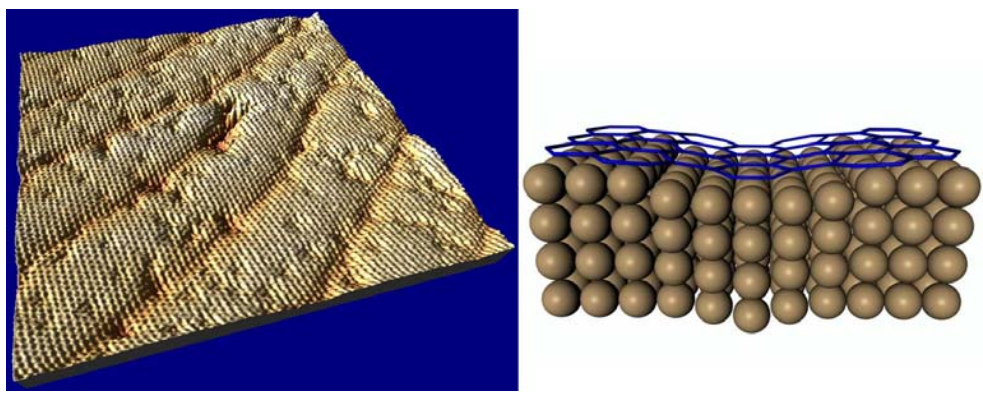

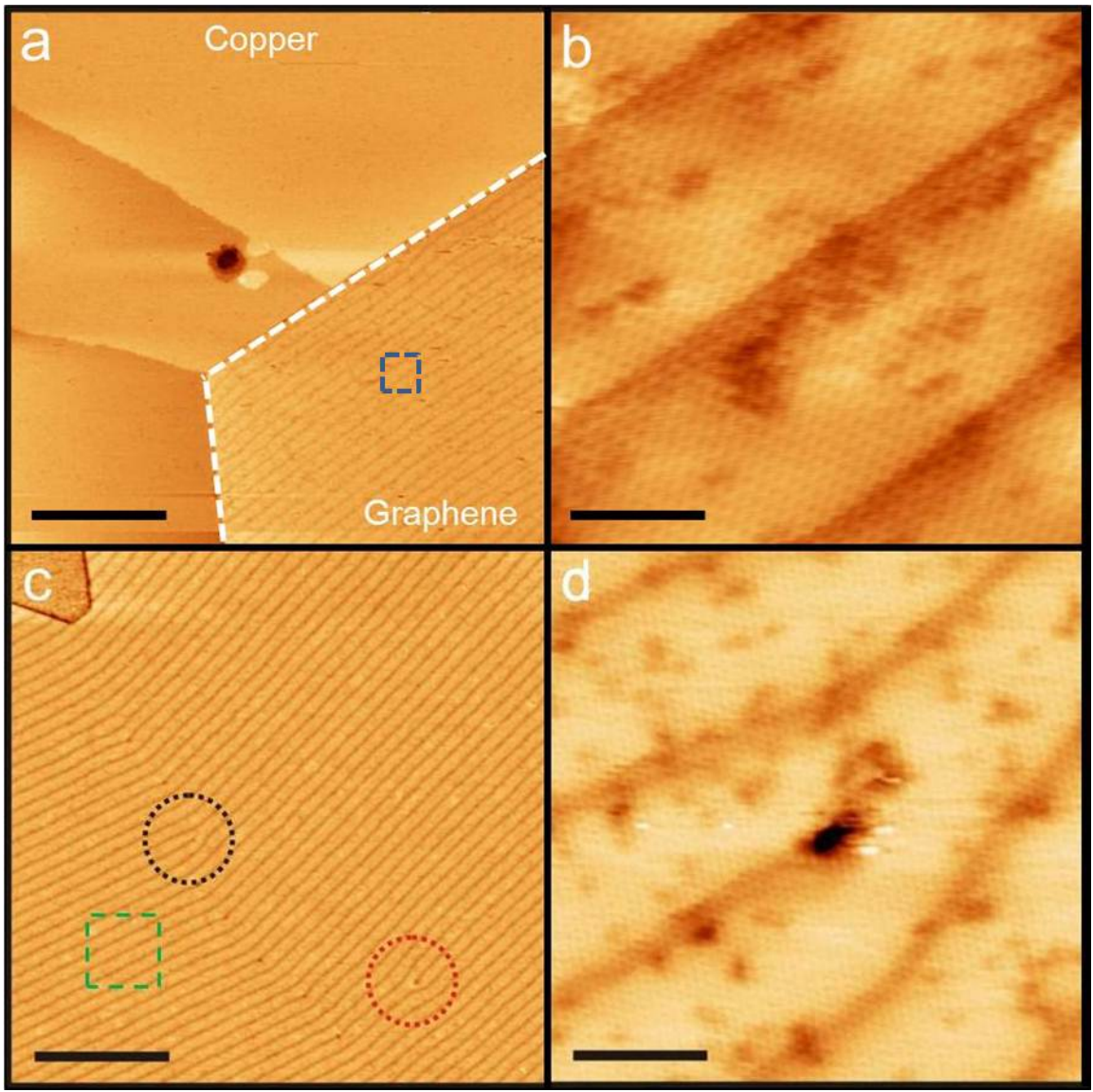

(Figure 1 by Tian *et al.*)

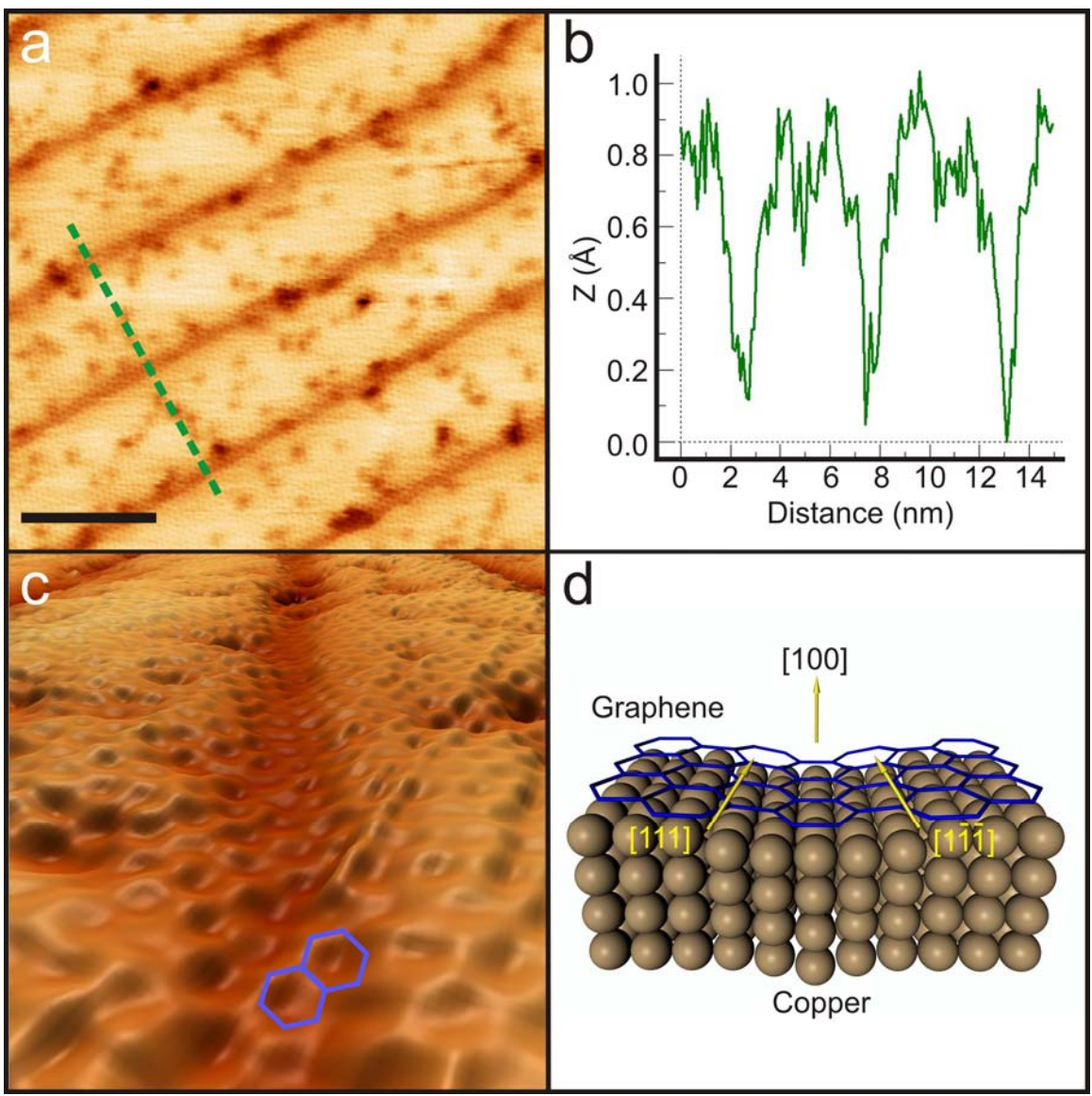

(Figure 2 by Tian *et al.*)

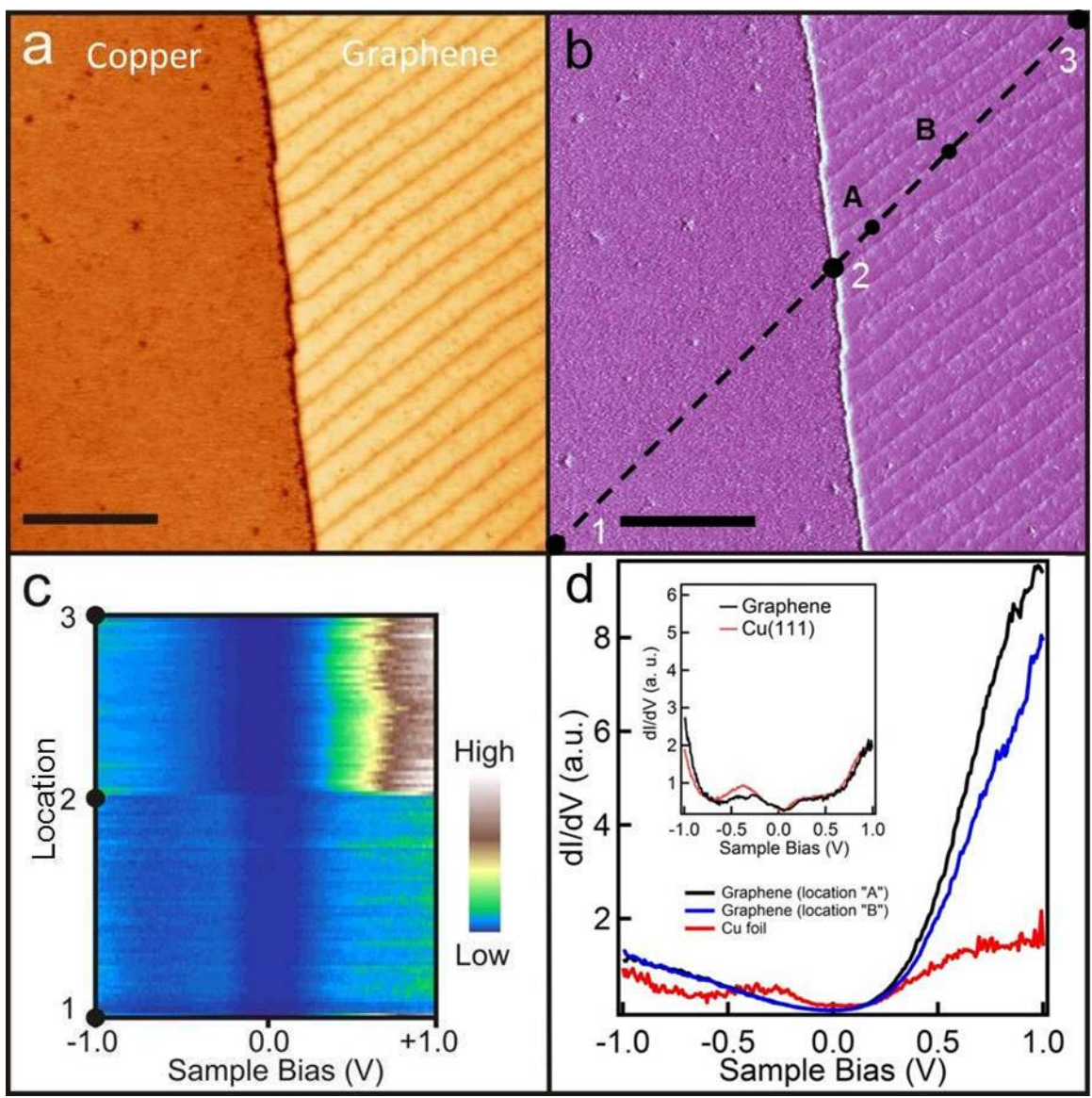

(Figure 3 by Tian *et al.*)